\documentclass[%
reprint,
aps,
superscriptaddress,
nofootinbib,
nobibnotes,
]{revtex4-1}

\usepackage{xcolor}
\usepackage{graphicx}
\graphicspath{{./fig/}{./fig/SolitonSpikeNFW_SIDM-BHL-Ostri/}}
\usepackage[export]{adjustbox}
\usepackage{amsmath,amssymb,amsfonts,mathrsfs,bm}
\usepackage[%
colorlinks=true,
linkcolor=blue,
citecolor=blue,
]{hyperref}

\newcommand{\eqcontrimark}{\star}
\newcommand{\eqcontri}{$^{\color{blue}\eqcontrimark}$}

\newcommand{\dif}{\mathrm{d}}

\newcommand{\Fig}{Fig.~}

\newcommand{\Msun}{M_{\odot}}
\newcommand{\yr}{\mathrm{yr}}

\newcommand{\Hz}{\mathrm{Hz}}
\newcommand{\mHz}{\mathrm{mHz}}

\newcommand{\LCDM}{$\Lambda$CDM }

\newcommand{\GW}{\mathrm{GW}}
\newcommand{\DF}{\mathrm{DF}}
\newcommand{\ac}{\mathrm{Ac}}
\newcommand{\sol}{\mathrm{soliton}}
\newcommand{\Spk}{\mathrm{spike}}
\newcommand{\spk}{\mathrm{sp}}
\newcommand{\NFW}{\mathrm{NFW}}

\newcommand{\ISCO}{\mathrm{ISCO}}
\newcommand{\AU}{\mathrm{AU}}

\newcommand{\halo}{\mathrm{halo}}

\newcommand{\rc}{r_{c}}

\newcommand{\m}[1]{m_{#1}}
\newcommand{\M}[1]{M_{#1}}

\newcommand{\omegas}{\omega_{\mathrm{s}}}

\newcommand{\Ncyc}{N_{\mathrm{cyc}}}

\newcommand{\rSch}{r_{\mathrm{S}}}
\newcommand{\Mc}{M_{c}}
\newcommand{\tret}{t_{\mathrm{ret}}}

\newcommand{\cs}{c_{s}}
\newcommand{\Mach}{\mathscr{M}}

\begin{document}

\title{%
Gravitational wave probes on self-interacting dark matter surrounding\\an intermediate mass black hole
}

\author{Kenji Kadota\eqcontri}
\email{kadota@ucas.ac.cn}
\affiliation{School of Fundamental Physics and Mathematical Sciences, Hangzhou Institute for Advanced Study, University of Chinese Academy of Sciences (HIAS-UCAS), Hangzhou 310024, China}
\affiliation{International Centre for Theoretical Physics Asia-Pacific (ICTP-AP), Beijing/Hangzhou, China}

\author{Jeong Han Kim\eqcontri}
\email{jeonghan.kim@cbu.ac.kr}
\affiliation{Department of Physics, Chungbuk National University, Cheongju, Chungbuk 28644, Korea}
\affiliation{School of Physics, Korea Institute for Advanced Study, Seoul 02455, Republic of Korea}

\author{Pyungwon Ko}
\email{pko@kias.re.kr}
\affiliation{School of Physics, Korea Institute for Advanced Study, Seoul 02455, Republic of Korea}
\affiliation{Quantum Universe Center (QUC), Korea Institute for Advanced Study, Seoul 02455, Republic of Korea}

\author{Xing-Yu Yang\eqcontri,}
\email[Corresponding author:~]{xingyuyang@kias.re.kr}
\affiliation{Quantum Universe Center (QUC), Korea Institute for Advanced Study, Seoul 02455, Republic of Korea}

\def\thefootnote{$\eqcontrimark$}
\footnotetext{These authors contributed equally to this work.}
\def\thefootnote{\arabic{footnote}}
\setcounter{footnote}{0}

\begin{abstract}
The presence of dark matter overdensities surrounding a black hole can influence the evolution of a binary system.
The gravitational wave signals emitted by a black hole binary offer a promising means to probe the dark matter environments near a black hole.
The dense region of dark matter can lead to the dephasing of gravitational waveforms, which can be detected by upcoming experiments such as the Laser Interferometer Space Antenna (LISA).
The dark matter density profile around the black hole can vary for different dark matter models.
Our study specifically investigates the impact of the ultralight self-interacting scalar dark matter (SIDM) on the gravitational wave signals emitted by black hole binaries.
A distinctive characteristic of SIDM surrounding a black hole, as opposed to collisionless dark matter, is the formation of a soliton core.
We perform a Fisher matrix analysis to estimate the size of the soliton and the corresponding SIDM parameter space that future LISA-like gravitational wave experiments can explore.
\end{abstract}

\maketitle

\section{Introduction}

A wealth of compelling experimental data, such as the galaxy rotation curves and gravitational lensing observations, firmly supports the existence of dark matter (DM) \cite{1980ApJ...238..471R,Corbelli:1999af, Refregier:2003ct, Allen:2011zs, Rubakov:2019lyf}.
In pursuit of unraveling the nature of dark matter, extensive attempts have been undertaken including  direct and indirect searches \cite{Graham:2015ouw, Schumann:2019eaa, Gaskins:2016cha} as well as collider studies \cite{Boveia:2018yeb}.
However, despite these significant efforts, precise properties of DM such as its mass and interactions remain largely unknown.
This enduring challenge of understanding DM stands as one of the longstanding conundrums in modern physics \cite{Bertone:2016nfn}.
Even though the cold dark matter (CDM), which serves as a fundamental component in the highly successful \LCDM model, has played a remarkable role in elucidating a large-scale structure of the Universe \cite{Planck:2018vyg}, it encounters challenges such as the too-big-to-fail problem and the core-cusp problem when scrutinizing smaller scales $\lesssim 1~\text{kpc}$ \cite{Boylan-Kolchin:2011qkt,Garrison-Kimmel:2014vqa,Papastergis:2014aba,deBlok:1997zlw,vandenBosch:1999ka,deBlok:2001hbg,KuziodeNaray:2007qi,Klypin:1999uc,Moore:1999nt}.
The null results from DM search experiments, combined with the presence of those unsettled small-scale structure issues, have spurred investigations into alternative avenues beyond the conventional CDM paradigm.

We study one such possible DM candidate, the ultralight self-interacting scalar dark matter, which is dubbed as SIDM for brevity~\cite{Lee:1995af,Spergel:1999mh, Peebles:2000yy, Goodman:2000tg, Arbey:2003sj,  Guzman:2006yc, Chavanis:2011zi, Chavanis:2011zm, Suarez:2011yf, Harko:2011jy, Guth:2014hsa, Eby:2015hsq, Marsh:2015wka, Fan:2016rda,Lee:2017qve,Chavanis:2018pkx, Eby:2018ufi,  Amin:2019ums, Garcia:2023abs,  Brax:2019fzb, Brax:2019npi, Boudon:2022dxi,Chakrabarti:2022owq,Dave:2023wjq}.
A prominent characteristic of such SIDM is a capability to form a stable configuration consisting of condensed bosonic fields, and the presence of such soliton cores can alleviate the tensions on small-scale structures.
By analyzing observations which can be affected by the existence of such a soliton, one can expect to extract valuable information about its fundamental properties.
As a concrete example for such astrophysical observations, we explore a black hole binary system in the presence of SIDM and corresponding gravitational wave (GW) signals.
Our goal is to investigate how future GW satellite experiments can be utilized to probe the properties of SIDM such as its mass and coupling.

For concreteness, our discussions focus on an intermediate mass black hole (IMBH) (typical mass range of order $\sim 10^2 M_{\odot}- 10^5M_{\odot}$ which is the mass situated between stellar-mass black holes and supermassive black holes).
The DM concentration in a vicinity of the IMBH is often referred to as ``minispike" in contrast to the DM ``spike'' profile around a supermassive black hole (SMBH) at the center of a galaxy.
IMBHs could form, for instance, in the centers of dwarf galaxies and globular clusters, and many IMBHs can well exist inside a single galaxy \cite{Bertone:2005xz,Zhao:2005zr}.
Compared to SMBHs at the centers of galaxies, IMBHs are less likely to experience major mergers because of their association to smaller halos and galaxy masses, and yet surrounding DM profiles can be sufficiently dense to withstand tidal disruptions.
The minispikes around IMBHs may hence well persist until the present epoch, which is a topic of active research.
As an observational probe of such DM environments, we estimate GW signals originating from a black hole binary involving an IMBH.
The GW frequency from the binary system depends on a mass ratio, and a stellar mass black hole orbiting the IMBH is of particular interest because the GW signals fall within the frequency band sensitive to planned space-based interferometers such as LISA ($\sim 0.1\mHz-1\Hz$)~\cite{LISA:2017pwj}.

The waveform of GWs is intricately linked to the properties of surrounding DM environments.
In our investigation, we specifically focus on studying the effects of SIDM surrounding an IMBH.
Notably, the presence of a solitonic core within a minispike marks this scenario as distinct from the vacuum case as well as from the conventional collisionless DM scenarios.
The dynamics of a stellar black hole spiraling through such a DM environment are influenced by a dynamical friction and an accretion.
These interactions result in accumulated phase shifts in gravitational waveforms, deviating from the predictions of the vacuum case.
Such effects on the gravitational waveforms have been investigated actively for the collisionless DM spike profile surrounding a central black hole which an accompanying black hole moves through \cite{Eda:2013gg,Eda:2014kra, Yue:2017iwc,Macedo:2013qea,Barausse:2014tra,Bertone:2019irm,Cole:2022fir,Kavanagh:2020cfn,Cardoso:2019rou,Hannuksela:2019vip,Coogan:2021uqv,Kim:2022mdj}.
More recently, the effect of a SIDM cloud has also been studied when a black hole binary system transverses through it, performing the waveform analysis in the time domain \cite{Boudon:2023vzl}.
Deviations in the speed of GWs that pass through the SIDM halo have also been studied to probe the mass and self-interaction~\cite{Dev:2016hxv}.
Further exploration and analysis of various DM models and configurations would provide an additional motivation for future GW experiments, enabling us to gain a deeper understanding of the nature of dark matter.

The aim of our study is to demonstrate that future GW experiments, such as LISA,
have the potential to identify SIDM through its distinctive stable self-bounded structure in the central region of a black hole binary system.
This study would contribute to shed light on the nature of dark matter and explore the unique signatures it may imprint on GW observations.

Our paper is structured as follows.
Section \ref{sec:soliton} overviews a soliton core formation around a black hole, which sets up the model to be investigated for the GW signal estimation.
Section \ref{sec:Inspiral} then describes the evolution of a binary system in the presence of the SIDM soliton core.
We discuss how a stellar black hole spiraling around a central IMBH can be affected by a dynamical friction and an accretion.
Section \ref{sec:fisher} presents our main results of the gravitational signal analysis.
We clarify the SIDM parameter space which the planned satellite experiments such as LISA can probe, along with the accuracy of the DM parameter estimation from a Fisher matrix analysis.
Section \ref{sec:dis} is devoted to a discussion/conclusion.

\section{Self-interacting dark matter around a Black Hole}\label{sec:soliton}

In this work, we consider a real scalar field $\phi$ SIDM model based on the following action:
\begin{equation}
    S=\int \dif^{4}x \sqrt{-g} \left[ -\frac{1}{2} g^{\mu\nu} \partial_{\mu}\phi \partial_{\nu}\phi - \frac{m^{2}}{2}\phi^{2} -\frac{\lambda}{4}\phi^{4} \right],
\label{eq:action}
\end{equation}
where $m$ is a mass of the scalar field, and $\lambda$ denotes a coupling constant responsible for a quartic self-interaction.
We assume that $\lambda$ is positive so that the interaction is repulsive which can counterbalance the gravity to form a soliton inside a halo.
The volume factor $\dif^{4}x \sqrt{-g}$ is invariant under a general coordinate transformation where $g = \text{det}(g_{\mu \nu})$ denotes a determinant of the Friedmann–Lema\^{i}tre–Robertson–Walker (FLRW) metric in Newtonian gauge
\begin{equation}
    ds^2 = (1 + 2 \Phi)dt^2 - (1 - 2 \Phi) a(t)^2 \delta_{ij} dx^i dx^j .
\label{eq:metric}
\end{equation}
Here, $a(t)$ is a cosmological scale factor, and $\Phi$ denotes a scalar perturbation neglecting the contribution of an anisotropic stress tensor.
We set $c=\hbar=1$ in this section for the notational brevity.

In the nonrelativistic limit where the momentum of particles is of negligible magnitude, $|\vec{p}| \ll m$,
we can rewrite the real scalar field $\phi$ in terms of a slowly varying complex field $\psi$ and a fast-varying phase $e^{\pm i mt}$:
\begin{equation}
    \phi=\frac{1}{\sqrt{2m}} \left( e^{-i mt}\psi + e^{i mt}\psi^{*} \right) .
\label{eq:phi}
\end{equation}
The time and momentum variances of $\psi$ are much smaller than $m$, i.e. $\dot{\psi}/\psi \ll m$ and $\nabla\psi/\psi \ll m$.

Substituting Eqs.~\eqref{eq:metric} and \eqref{eq:phi} into Eq.~(\ref{eq:action}), and neglecting fast oscillatory terms $e^{\pm 2imt}$ that average out to zero, we have
\begin{eqnarray} \nonumber
    S_{\phi} &=& \int \dif^{4}x~ a^3 \Big( \frac{i}{2} (\dot{\psi} \psi^* - \psi \dot{\psi}^* )
    - \frac{1}{2 m a^2} \partial_i \psi \partial^i \psi^* \\
    &-& m \Phi |\psi|^2  -  \frac{3 \lambda}{8 m^2} |\psi|^4 \Big),
\end{eqnarray}
where we have kept the terms up to the first order, and the superscript ``$\cdot$" denotes a differentiation with respect to time $t$.
The equation of motion for $\psi$ gives the nonlinear Schr\"{o}dinger equation in the FLRW spacetime,
\begin{equation}
    i \dot{\psi} =  -\frac{3}{2} H i \psi  -\frac{\nabla^{2}\psi}{2ma^2} + m \left( \Phi + \Phi_{\mathrm{\text{self}}} \right)\psi ,
\label{eq:Schrodinger1}
\end{equation}
where $H = \dot{a}/a $ denotes a Hubble parameter.
The nonlinear term arises from the self-interaction potential $\Phi_{\mathrm{\text{self}}}$ defined by
\begin{equation}
   \Phi_{\mathrm{\text{self}}} \equiv \frac{3 \lambda |\psi|^2}{4 m^3} \;.
\end{equation}
On galactic scales, we assume that the system of scalar fields has been completely decoupled from the background evolution of the Universe, in which case we can neglect the scale factor and the Hubble parameter to rewrite
Eq.~\eqref{eq:Schrodinger1} as
\begin{equation}
    i \dot{\psi} = -\frac{\nabla^{2}\psi}{2m} + m \left( \Phi + \Phi_{\mathrm{self}} \right)\psi \;.
\label{eq:Schrodinger2}
\end{equation}
When a high density medium of scalar fields condensates into the lowest momentum state,
it behaves like a single macroscopic fluid exhibiting a superfluidity.
The many-particle system of Eq.~(\ref{eq:Schrodinger2}) that involves a multitude of scalar fields is generically difficult to solve.
Therefore, we pursue the macroscopic analysis of the superfluid~\cite{GROSS195857,1961NCim...20..454G,Pitaevskii:1959, 2016arXiv160509580B, BE:2016}, by using a mean-field approximation
\begin{eqnarray}
    \psi ( \vec{r}, t ) = \hat{\psi}( \vec{r}, t ) + \delta \hat{\psi}( \vec{r}, t ),
\label{eq:meanfield}
\end{eqnarray}
where $\hat{\psi}( \vec{r}, t )$ denotes a wave function of the condensate, and $ \delta \hat{\psi}( \vec{r}, t )$
is a small perturbation of the system.
Then the density of the condensate is given by $n( \vec{r}, t ) =  |\hat{\psi}( \vec{r}, t )|^2$.
This approach leads to the reduction of the many-body problem to one single field $\hat{\psi}( \vec{r}, t )$ averaging out the effects of all other particles.
In this description, we can decompose the condensate wave function using the Madelung transformation~\cite{1927ZPhy...40..322M}
\begin{eqnarray} \nonumber
	\hat{\psi} &=& |\hat{\psi}( \vec{r}, t )| e^{i s(\vec{r}, t)} \\ \nonumber
	      &=& \sqrt{n( \vec{r}, t )} e^{i s(\vec{r}, t)}  \\
	      &=& \sqrt{\frac{\rho(\vec{r},t)}{m}} e^{i s(\vec{r}, t)},
\label{eq:Madelung}
\end{eqnarray}
where  $\rho(\vec{r},t) = n( \vec{r}, t ) m $ and $s$ denote an amplitude and a phase respectively.
Plugging Eqs.~\eqref{eq:meanfield} and \eqref{eq:Madelung} into Eq.~(\ref{eq:Schrodinger2}) and
taking imaginary and real parts result in hydrodynamical equations
\begin{eqnarray}
    \dot{\rho} + \nabla \cdot (\rho \vec{v}) &=& 0 \label{eq:hydro1}, \\
    \vec{v} + (\vec{v} \cdot \nabla) \vec{v} &=& - \nabla (\Phi + \Phi_{\mathrm{self}} + \Phi_{\mathrm{QP}}) \label{eq:hydro2} ,
\end{eqnarray}
where we have defined $\vec{v} = \frac{\nabla s}{m}$.
The term $\Phi_{\mathrm{QP}} = - \frac{\nabla^2 \sqrt{\rho}}{2 m^2 \sqrt{\rho}}$ represents a quantum pressure arising from the fact that a system cannot be of infinitesimal size due to the Heisenberg uncertainty principle.
We will work in the regime where the quantum pressure is negligibly smaller than the self-interaction potential
\begin{equation}
	\Phi_{\mathrm{self}}=  \frac{3 \lambda \rho}{4 m^4}  \gg \Phi_{\mathrm{QP}}.
\end{equation}

We consider a halo system comprised of a black hole (BH) of mass $M_{\text{BH}}$ at the center surrounded by the
condensate of scalar fields (i.e. dressed BH).
In the Newtonian limit, the metric perturbation can be written as
\begin{equation}
	\Phi = \Phi_{\text{BH}} +\Phi_{\text{dress}},
\end{equation}
where $\Phi_{\text{BH}} = - \frac{\rSch}{2r}$ denotes a BH potential with Schwarzschild radius $\rSch = 2 G M_{\text{BH}}$, and $\Phi_{\text{dress}}$ represents a potential generated by a gravitational interaction of the scalar fields
satisfying a Poisson equation
\begin{eqnarray}
	 \nabla^2 \Phi_{\text{dress}} &=& 4 \pi G \rho.
\end{eqnarray}

When a repulsive pressure due to the self-interaction balances out a gravity,
the system can form a stable configuration with vanishing velocity $\vec{v} = 0$,
in which case Eq.~(\ref{eq:hydro2}) reduces to
\begin{eqnarray}
	 \nabla ( \Phi+ \Phi_{\mathrm{self}} ) = 0 .
\label{eq:static}
\end{eqnarray}
Taking a divergence on both sides and assuming a spherically symmetric solution for $\rho$, we find
\begin{eqnarray}
	\frac{d^2 \rho}{d r^2} + \frac{2}{r} \frac{d \rho}{d r} + \frac{1}{r^2_c} \rho = 0,
\end{eqnarray}
where $\rc$ denotes a characteristic radius of a soliton defined by
\begin{eqnarray}\label{eq:rc}
	\rc \equiv \sqrt{\frac{3 \lambda}{16 \pi G m^4}}.
\end{eqnarray}
Changing a variable $x = r/\rc$ gives a spherical Bessel equation of order $\ell = 0$,
\begin{eqnarray}
	\frac{d^2 \rho}{d x^2} + \frac{2}{x} \frac{d \rho}{d x} + \rho = 0 ,
\end{eqnarray}
where there are two independent solutions, $j_{\ell=0} = \sin x / x$ and $n_{\ell =0} = \cos x /x$.
The final density profile of a soliton is given by a linear combination of these two solutions~\cite{Arbey:2003sj,Boehmer:2007um,Chavanis:2011zi,Brax:2019npi}
\begin{equation}
    \rho_\sol(r)=\rho_{\sin} \frac{\sin(r/\rc)}{r/\rc}+\rho_{\cos} \frac{\cos(r/\rc)}{r/\rc},
\label{eq:Soliton}
\end{equation}
where $\rho_{\sin}$ and $\rho_{\cos}$ are constants to be determined.
The second term dominates for a small $r$ where the gravitational pull from the black hole is significant.

Outside of a soliton ($r> \rc$), on the other hand, the scalar fields start to behave like the collisionless CDM.
To facilitate a smooth transition of a density profile from the outer to the inner region, we will use
a continuity condition and a mass conservation to match profiles in adjoint sectors.

First, in the outermost region of halo, we consider the Navarro-Frenk-White (NFW) density profile~\cite{Maccio:2008pcd}:
\begin{equation}
    \rho_{\NFW}(r)=\frac{\rho_{0}}{(r/r_{sc}) \left(1+r/r_{sc}\right)^{2}} \;,
\label{eq:NFW}
\end{equation}
\begin{equation}
    \rho_0=
    \frac{\rho_{\text{crit}} \Delta }{3}
    \frac{\mathfrak{c}^3}{\ln(1+\mathfrak{c})-\mathfrak{c}/(1+\mathfrak{c})},
\end{equation}
where the concentration parameter $\mathfrak{c}\equiv r_{\mathrm{vir}}/r_{sc}$ is a ratio between a virial radius and a scale radius,
and the halo density is $\rho_{\text{crit}} \Delta $ at the virial radius.
We take $\Delta=200$ and adopt the scaling relation between the concentration and the halo mass ($\mathfrak{c}-M$ relation) given by Ref. \cite{Correa:2015dva} for concreteness.
The critical density of the Universe today $\rho_{\text{crit}} = 2.775 h^2 \times 10^{11} M_{\odot}$ and the reduced Hubble constant $h = H_0/(100 \text{ km/s/Mpc})$ with $H_0 = 67.4
\text{ km/s/Mpc}$ inferred from the CMB~\cite{Planck:2018vyg}.

Second, as they approach closer to the center, due to the influence of the central BH, the scalar fields are redistributed to form a spike profile~\cite{Gondolo:1999ef,Ullio:2001fb}
\begin{equation}
\rho_{\Spk}(r)=\rho_{\spk} \left( \frac{r_{\spk}}{r} \right)^{\gamma_{\spk}},
\label{eq:spike}
\end{equation}
where $\rho_{\spk}$ denotes a density at a reference radius $r_{\spk}$, and
 $\gamma_{\spk}$ is a slope of the spike.
 Assuming an adiabatic growth of the spike in the presence of the central BH with an initial slope of $\gamma_i$,
 it acquires a final slope of $\gamma_{\spk} = (9-2\gamma_i)/(4-\gamma_i)$.
 For instance, $\gamma_i=1$ for the NFW and then $\gamma_{\spk}= 7/3$.
 The two undetermined parameters $\rho_{\spk}$ and $r_{\spk}$ are obtained by a continuity condition and a mass conservation
\begin{eqnarray}
	\rho_{\NFW}(r_{\spk}) &=& \rho_{\Spk}(r_{\spk})  , \\
	 \int_{r_{\min}}^{5 r_{\spk}} \rho_{\text{DM}}(r) 4\pi r^{2} \dif r &=& 2 M_{\text{BH}} ,
\end{eqnarray}
where $r_{\min}= 3 \rSch$ denotes an innermost stable circular orbit (ISCO), and
the upper limit of the integral $5 r_{\spk}$ is empirically obtained such that the resulting integral gives twice the mass of a BH~\cite{Eda:2014kra}.%
\footnote{%
For a standard collisionless cold DM around a BH, the DM would not maintain a stable orbit within $r_{\min}$  (e.g. spiral into the BH) and the DM density can vanish inside $r\lesssim 3 \rSch$.
Taking account of the relativistic effects, it may be more appropriate to set $\rho=0$ for $r\lesssim 2 \rSch$ \cite{Sadeghian:2013laa,Feng:2021qkj,DeLuca:2023laa}.
Moreover, a repulsive self-interaction in the scalar field may make $r_{\min}$ closer to the BH due to the additional outward force it provides, though it still should be bigger than the event horizon scale $r_{\min}>\rSch$.
These alterations in the factor of a few for $r_{\min}$ however do not have a noticeable impact on our results for the parameter accuracy estimation from expected gravitational wave signals.
We hence simply take $\rho=0$ for $r< 3 \rSch$ for concreteness in our discussions.
}
The integrand $\rho_{\text{DM}}$ is defined by
\begin{equation}
    \rho_{\text{DM}}(r)=
    \begin{cases}
       \rho_{\Spk}(r),   & r_{\min} \le r < r_{\spk}, \\
       \rho_{\NFW}(r),   & r_{\spk} \le r.  \\
    \end{cases}
\end{equation}

Third, since the innermost region of halo is described by the soliton configuration in Eq.~(\ref{eq:Soliton}),
we again use the continuity condition and the mass conservation to fix the parameters $\rho_{\sin}$ and $\rho_{\cos}$,
\begin{eqnarray}
	\rho_{\sol}(\rc) &=& \rho_{\Spk}(\rc)  \;, \\
	 \int_{r_{\min}}^{\rc} \rho_{\sol}(r) 4\pi r^{2} \dif r &=& \int_{r_{\min}}^{\rc} \rho_{\Spk}(r) 4\pi r^{2} \dif r .
\end{eqnarray}

\begin{figure}[htbp]
    \centering
    \includegraphics[]{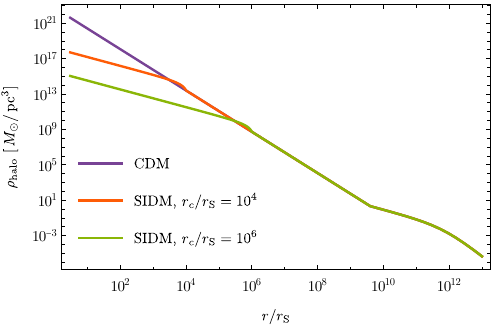}
    \caption{The DM density profile around a $10^{4}\Msun$ black hole in the CDM model and SIDM model with different soliton characteristic radii.
    }
    \label{fig:rhoHalo}
\end{figure}

To sum up, the final density profile of a halo is composed of three layers
\begin{equation}
    \rho_{\halo}(r)=
    \begin{cases}
       \rho_{\sol}(r),   & r_{\min} \le r < \rc, \\
       \rho_{\Spk}(r),   & \rc \le r < r_{\spk}, \\
       \rho_{\NFW}(r),   & r_{\spk} \le r.
    \end{cases}
\label{eq:profile}
\end{equation}
Figure~\ref{fig:rhoHalo} shows the density profile of a halo around a $10^{4}\Msun$ BH in the CDM model and SIDM model with different characteristic soliton radii for illustration.
One can find that densities of soliton cores in the SIDM model are smaller than the density in the CDM model.
This is because a greater repulsive self-interaction leads to a larger soliton core $r_c$ which reduces densities of solitons.

In Sec.~\ref{sec:Inspiral}, we will utilize this profile
to compute a dynamical friction and an accretion rate for a stellar-mass BH moving through an IMBH minispike.

In addition to the solitonic profile, another characteristic feature of SIDM to be compared with the collisionless DM is the effective sound speed due to the pressure caused by the self-interaction.
Our analysis employs an effective model where the SIDM system reacts to a moving object by generating phononlike sound waves as follows.

Perturbations of the condensate are manifested by the excitation in the SIDM superfluid.
They stand for sound waves that travel through the condensate.
To obtain a sound speed at which it travels~\cite{2016arXiv160509580B, BE:2016}, we substitute Eq.~(\ref{eq:meanfield}) into Eq.~(\ref{eq:Schrodinger2}) with neglecting the contributions of the gravity and the perturbation
\begin{equation}
    i \dot{\hat{\psi}} = -\frac{\nabla^{2}\hat{\psi}}{2m} + g n \hat{\psi} \;,
\label{eq:Schrodinger3}
\end{equation}
where $g = \frac{3 \lambda}{4 m^2}$.
The stationary solution which represents a ground state of the condensate takes the following form:
\begin{equation}
    \hat{\psi}(\vec{r},t) = \hat{\psi}_s(\vec{r}) e^{-i\mu t},
\label{eq:stationary}
\end{equation}
where $\mu$ is a coefficient to be determined.
Plugging this ansatz into Eq.~~(\ref{eq:Schrodinger3}) and neglecting the kinetic term give
\begin{equation}
    \mu = \frac{3 \lambda}{4m^2} n .
\label{eq:mu}
\end{equation}
The excited states can be found from seeking a solution of the form
\begin{equation}
    \hat{\psi}(\vec{r},t) = \big( \hat{\psi}_s (\vec{r}) + \delta  \hat{\psi}_s (\vec{r},t)  \big) e^{-i\mu t},
\label{eq:Excited}
\end{equation}
where
\begin{equation}
     \delta  \hat{\psi}_s (\vec{r},t) = u(\vec{r}) e^{-iwt} - v^*(\vec{r}) e^{iwt}.
\label{eq:Excited2}
\end{equation}
Plugging this ansatz into Eq.~(\ref{eq:Schrodinger3}), and taking the Fourier transformation give
\begin{eqnarray}
    \Big( \frac{k^2}{2m} - w + g n \Big) \hat{u}(\vec{k})- g n \hat{v}(\vec{k}) = 0  ,\\
    \Big( \frac{k^2}{2m} + w + g n \Big) \hat{v}(\vec{k}) - g n \hat{u}(\vec{k}) = 0 .
\end{eqnarray}
Combining these equations gives the dispersion relation in $k$-space,
\begin{equation}
     w^2 = \frac{k^4}{(2m)^2} + c^2_s k^2 ,
\label{eq:dispersion}
\end{equation}
where the sound speed is defined by
\begin{equation}
     c^2_s \equiv \frac{3 \lambda}{4 m^3}n = \frac{3 \lambda \rho }{4 m^4} .
\label{eq:soundspeed}
\end{equation}
The sound speed originates from the self-interaction, and it changes the energy spectrum of exited states.

In the small $k$ limit (long-range), the first term in Eq.~(\ref{eq:dispersion}) dominates
\begin{equation}
     w \simeq c_s k = \frac{3 \lambda \rho }{4 m^4} k .
\label{eq:dispersion2}
\end{equation}
This implies that the repulsive interaction ($\lambda > 0$) results in an oscillating solution.
However, the attractive interaction ($\lambda < 0$) gives either a growing or a decaying mode, indicating that it cannot form a stable condensate\footnote{See Refs.~\cite{Guzman:2006yc,Amin:2019ums} for the case of the attractive interaction to form a stable soliton.}.

We also utilize this derived sound speed in Sec.~\ref{sec:Inspiral} in studying the evolution of stellar-mass BHs traversing through a soliton core.

\section{Black hole binary evolution}\label{sec:Inspiral}

In the previous section, we have discussed the DM halo which surrounds a BH with a mass $\m1 (\equiv M_{\text{BH}})$.
When a stellar-mass BH with a mass $\m2$ is caught gravitationally by such a BH, the formed binary system can produce the observable GWs.
The GWs from such a binary system carry the information of a surrounding DM halo; thus one can probe the property of the DM environment with these GWs~\cite{Eda:2013gg,Eda:2014kra}.
In this work, we focus on the case with $q \equiv \m2/\m1 \ll 1$ which can emit GWs in the frequency range of LISA.

With the density profile of the DM halo, one can derive dynamical equations of the binary system.
The equation of motion in a radial direction is
\begin{equation}
    \dot{m}_{2} \dot{r}_{2} - \m2 r_{2} \omegas^{2} + \m2 \ddot{r}_{2} = - G \frac{\M1 \m2}{r^{2}},
\end{equation}
where $r_{2}$ is the distance between a stellar-mass BH and the barycenter of the system, $r$ is the relative distance of two objects, $\omegas$ is the angular velocity, and $\M1$ is the total mass inside the orbit given by
\begin{equation}\label{eq:M1}
    \M1=
    \begin{cases}
      \m1,   & r <  r_{\min}, \\
      \m1 + \int_{r_{\min}}^{r} \rho_{\halo}(r')4\pi r'^{2} \dif r',  & r_{\min} \le r,
    \end{cases}
\end{equation}
which includes the contributions from the central BH and the surrounding DM halo.

Since the mass ratio $q$ is much less than $1$, the orbit can be regarded as a quasicircular orbit and the DM halo can be considered as unperturbed.\footnote{Refs.~\cite{Kavanagh:2020cfn,Coogan:2021uqv} discussed the halo feedback in a similar setup but focused on the CDM case. }
In this approximation, both $\dot{r}_{2}$ and $\ddot{r}_{2}$ vanish, and we obtain
\begin{equation}\label{eq:omegas}
    \omegas=\sqrt{\frac{G M}{r^{3}}},
\end{equation}
where $M\equiv \M1+\m2$ is a total mass of the system.

In addition to the gravitational pull of DM inside the orbit, the DM halo can also affect the evolution of the binary system by a dynamical friction and an accretion.
Since we consider a small stellar-mass BH orbiting a large central BH with its own unperturbed DM halo, only the dynamical friction and the accretion affecting the dynamics of the stellar-mass BH are taken into account.

The combined effects of GW emissions, the dynamical friction, and the accretion decrease the orbital energy and reduce the relative distance of the binary.
The evolution of relative distance reads as
\begin{equation}
    \dot{r} = - \left( F_{\GW} +F_{\DF} + F_{\ac} \right) \left( 2\mu\omegas + \mu r \frac{\dif \omegas}{\dif r} \right)^{-1},
\end{equation}
where $\mu\equiv \M1\m2/M$ denotes a reduced mass, and the backreaction force due to the emission of GWs is given by~\cite{Maggiore:2007ulw}
\begin{equation}\label{eq:FGW}
    F_{\GW}=\frac{1}{v}\frac{32 G^{4} \mu^{2} M^{3}}{5 c^{5} r^{5}},
\end{equation}
where $v\equiv r \omegas$ is the velocity of a stellar-mass BH.

The dynamical friction of a stellar-mass BH orbiting inside the SIDM medium takes the form~\cite{Ostriker:1998fa}
\begin{equation}
    F_{\DF} =\frac{4\pi (G \m2)^{2} \rho_{\halo}}{v^{2}} I(\Mach, \Lambda),
\label{eq:DF_SIDM}
\end{equation}
where the function $I$ is defined by
\begin{equation}
    I(\Mach, \Lambda)=
     \begin{cases}
      \frac{1}{2}\ln \left( \frac{1+\Mach}{1-\Mach} \right) -\Mach,   & \Mach<1, \\
     \frac{1}{2}\ln \left( 1-\Mach^{-2} \right) + \ln \Lambda ,   & \Mach>1,
    \end{cases}
    \label{eq:I}
\end{equation}
where $\Lambda\equiv vt/r_{\min}$ with $t$ denoting the time for which the stellar-mass BH has traveled.
The Coulomb logarithm $\ln \Lambda$ incorporates the minimum and maximum impact parameters of the stellar mass BH.
The Mach number $\Mach\equiv v/\cs$ parametrizes how fast the stellar-mass BH moves through the medium
with respect to the sound speed $c_s$ defined by Eq.~(\ref{eq:soundspeed})
\begin{equation}
c^2_s = \frac{3 \lambda \rho_{\text{halo}}}{4 m^4} \;,
\end{equation}
with substituting the density profile with Eq.~(\ref{eq:profile}).
For the supersonic expression, in the limit $\Mach \gg 1$, we recover the conventional steady state collisionless dark matter result
$I\rightarrow \ln (vt/r_{\min})$ with the replacement of $vt\rightarrow r_{\max}$.
Here we use the Coulomb logarithm $ \ln \Lambda= \ln \sqrt{m_1/m_2}$ which is commonly adopted in the literature~\cite{ Kavanagh:2020cfn,2008gady.book.....B, 2010gfe..book.....M}\footnote{Such a choice of $\ln \Lambda$ would suffice for our purpose of demonstrating the potential gravitational signals, partly because of its weak logarithmic dependence and also other relevant uncertainties such as the modeling of GW detector noise and sensitivity.
The exact value of the Coulomb term $\Lambda$ is not well known and is usually obtained numerically in the literature.
For instance, $r_{\min}$ can be estimated by demanding that the numerically calculated force should match the analytically estimated one \cite{Morton:2021qbo}.
The value of $r_{\max}$ can be, for instance, the radius under the influence of BH gravity (Roche radius or Hill radius) or the soliton radius.
Different authors use different values for the Coulomb term, and we refer the readers to the existing literature (e.g. \cite{Ostriker:1998fa,Berezhiani:2019pzd,Morton:2021qbo,Kavanagh:2020cfn,Boudon:2022dxi,Boudon:2023vzl}) for more detailed discussions on the dynamical friction in the presence of the gaseous medium and collisional DM.}.
In the subsonic limit $\Mach \ll 1$, on the other hand, the dynamical friction exerted by the collisionless medium is larger
than the SIDM partly due to the self-interacting repulsive force that hinders the accumulation of the DM in the wake.

The $F_{\ac}$ denotes the accretion drag force
\begin{equation}
    F_{\ac}=\dot{\mu} v,
\end{equation}
where $\dot{\mu}=\dot{m}_{2}(1+q)^{-2}$, and $\dot{m}_{2}$ denotes the accretion rate.
For the SIDM medium, we adopt the
Bondi-Hoyle-Lyttleton accretion rate~\cite{1939PCPS...35..405H,1952MNRAS.112..195B,1944MNRAS.104..273B,1983bhwd.book.....S,Macedo:2013qea}
\begin{equation}\label{eq:acc-SIDM}
    \dot{m}_{2} = \frac{4\pi (G \m2)^{2} \rho_{\halo}}{(\cs^{2}+v^{2})^{3/2}},
\end{equation}
while for the collisionless medium we use the form
\begin{equation}
    \dot{m}_{2} = \frac{16\pi (G \m2)^{2} \rho_{\halo}}{c^2 v} \big( 1 + \frac{v^2}{c^2}\big).
\end{equation}
In the transonic regime $v \sim c_s\ll c$, for example, the accretion rate in the SIDM medium is much larger than the collisionless case
\begin{equation}
    \frac{\dot{m}_{2,\text{CDM}}}{\dot{m}_{2,\text{SIDM}}} \sim \frac{v^2}{c^2} \ll 1.
\end{equation}
Such an enhancement can arise from ``funneling effect", in which the particles tend to funnel into inward-spiraling trajectories \cite{1983bhwd.book.....S}.
The collisions and interactions among the particles redistribute their angular momenta, and, combined with the gravitational influence of the BH, the particles tend to follow more radial trajectories toward the accreting BH.

The relative magnitude of the accretion drag force and the dynamical friction for the collisionless medium is
\begin{equation}
    \frac{F_{\DF,\text{CDM}}}{F_{\ac,\text{CDM}}} \sim \frac{c^2}{v^2} \gg 1 ,
\end{equation}
and for the SIDM case, in the supersonic limit $v \gg c_s$,
\begin{equation}
    \frac{F_{\DF,\text{SIDM}}}{F_{\ac,\text{SIDM}}} \gtrsim 1 .
\end{equation}

In comparison to the binary in the vacuum, the binary in the DM halo can be also affected by the dynamical friction and accretion.
These effects accelerate the loss of orbit energy of the binary; thus the number of orbital cycles before the coalescence $\Ncyc$ gets smaller.

\begin{figure}[htbp]
    \centering
    \includegraphics[]{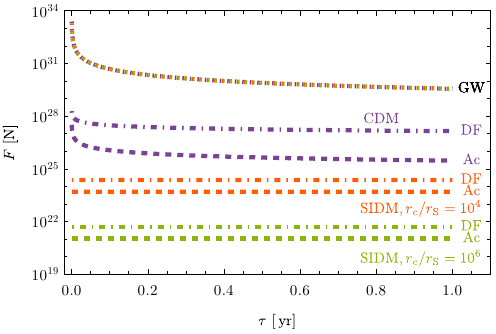}
    \caption{The evolution of the forces contributed by GWs, dynamical frictions, and accretions in the CDM model and the SIDM model with different soliton characteristic radii.
    $\tau$ is the time to reach the ISCO.
        The force due to the gravitational wave emission is dominant, and three GW curves are indistinguishable among different parameter sets in this figure (represented by the top dotted curve).
}
    \label{fig:F}
\end{figure}

In \Fig\ref{fig:F}, we show the evolution of forces contributed by GWs, dynamical frictions, and accretions respectively where $\tau \equiv t|_{r=r_{\ISCO}} -t$ is the time to ISCO, and benchmark parameters of the binary system are given by the set 1 in Table~\ref{tab:paras}.
One can find that, in the final stage of inspiral which is relevant for  GW observations, the force due to the emission of GWs is dominant for both CDM and SIDM halos; thus the orbital decay of the binary is still mainly determined by the emission of GWs, rather than by the dynamical friction or accretion.

\begin{table}[htbp]
    \centering
    \caption{
    The benchmark parameter sets we use in the analysis.
    The mass $m_2$ changes with time due to the accretion, and the values here are given when $r=r_{\ISCO}$.
}
    \label{tab:paras}
    \begin{tabular}{|c|c|c|c|}
        \hline
        Set & $\m2/\Msun$ & $\m1/\Msun$ &  $M_{200}/\Msun$  \\
        \hline
        1 & 1 & $10^{4}$ & $10^{8}$  \\
        \hline
        2 & 10 & $10^{4}$ & $10^{8}$  \\
        \hline
        3 & 10 & $10^{5}$ & $10^{10}$  \\
        \hline
    \end{tabular}
\end{table}

\begin{figure}[htbp]
    \centering
    \includegraphics[]{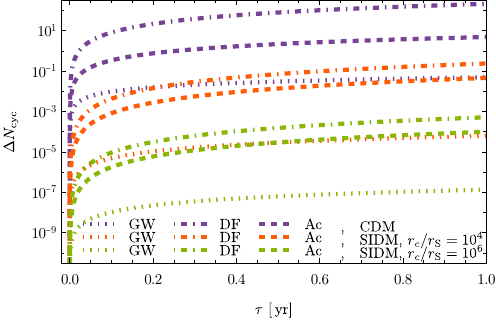}
    \caption{The difference in the number of orbital cycles compared with the binary in the vacuum,
    $\Delta \Ncyc=\Ncyc\text{(vacuum)}-\Ncyc\text{(with DM halo)}$, as a function of the time $\tau$ to reach the ISCO.
 }
    \label{fig:DeltaNcycFall}
\end{figure}

Figure~\ref{fig:DeltaNcycFall} demonstrates the corresponding difference of $\Ncyc$ compared to the binary in the vacuum, $\Delta \Ncyc=\Ncyc\text{(vacuum)}-\Ncyc\text{(with DM halo)}$.
One can find that $\Delta \Ncyc$ is primarily contributed by the dynamical friction (represented by dot-dashed lines), while the contribution from the accretion (dashed lines) is subdominant.
Even though the force due to the GW emission is significant and hence can affect how quickly the stellar-mass BH approaches toward the central BH, its contribution to the accumulated phased shift in the waveform is negligible.

\section{Gravitational wave analysis}\label{sec:fisher}

The waveform of GWs produced from the inspiral of a binary system is given by~\cite{Maggiore:2007ulw}
\begin{align} \nonumber
    h_{+}(t) =&\frac{4}{D_{L}} \left( \frac{G\Mc}{c^{2}} \right)^{5/3} \left[ \frac{\pi f(\tret)}{c} \right]^{2/3}\frac{(1+\cos^{2}\iota)}{2} \\
    &\times \cos[\Psi(\tret)],\\ \nonumber
    h_{\times}(t)=&\frac{4}{D_{L}} \left( \frac{G\Mc}{c^{2}} \right)^{5/3} \left[ \frac{\pi f(\tret)}{c} \right]^{2/3} \cos\iota\ \\
    &\times \sin[\Psi(\tret)],
\end{align}
where $D_{L}$ is a luminosity distance to a source, $\Mc=\mu^{3/5}M^{2/5}$ denotes a chirp mass, $\tret=t-D_{L}/c$ is a retarded time, $\iota$ is the angle between an orbital angular momentum axis of a binary and a direction to a detector, $f$ is a frequency of GWs which is given by $f=\omega_{\GW}/(2\pi)$ with $\omega_{\GW}=2\omegas$, and $\Psi$ is a phase of GWs.

\begin{figure}[htbp]
    \centering
    \includegraphics[]{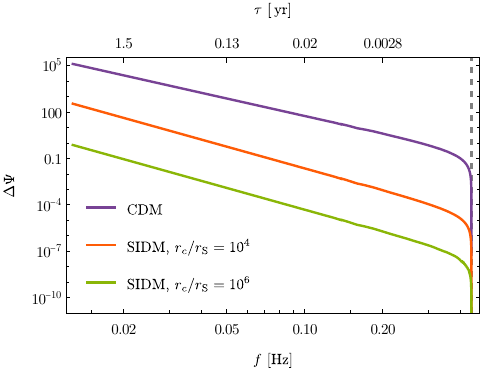}
    \caption{The dephasing of the gravitational waveform for our example parameter sets with respect to the vacuum case, $\Delta \Psi = \Psi\text{(vacuum)}-\Psi\text{(with DM halo)}$ as a function of the GW frequency.
    The vertical dashed line denotes the frequency at the ISCO.
    The corresponding $\tau$ (the time to reach the ISCO) at each frequency is also shown.
    }
    \label{fig:dephasing}
\end{figure}

Figure~\ref{fig:dephasing} shows the difference in the phase of GWs,
\begin{equation}
\Delta \Psi = \Psi\text{(vacuum)}-\Psi\text{(with DM halo)}
\end{equation}
for different DM parameter sets compared with the vacuum case, and the parameters of the binary system are given by the set 1 in Table~\ref{tab:paras}.
One can find the phase of the GW waveform is affected, the so-called dephasing effect, by the DM environment.
Over the inspiral period, this phase shift effect can accumulate and result in observable signals of sufficient magnitude to probe the characteristics of the dark matter environment.

Even though the dephasing effect is calculated with a numerical approach in our quantitative discussions, it would be informative to provide an analytical assessment to see how it is affected by our model parameters.
The conservation of energy gives
\begin{equation}\label{eq:eng_con}
    -\frac{\dif E_{\mathrm{orb}}}{\dif t} = P_\GW + P_\DF + P_{\ac},
\end{equation}
where the orbital energy of the binary is
\begin{equation}
    E_{\mathrm{orb}}=-(G^2 \Mc^5 \omega_\GW^2/32)^{1/3},
\end{equation}
and the power of energy loss due to the GW emission, the dynamical friction, and the accretion are respectively given by
\begin{align}
    P_\GW &=\frac{32}{5}\frac{c^5}{G}(\frac{G\Mc\omega_\GW}{2c^3})^{10/3}, \\
    P_{\DF} &=v\,F_{\DF} , \\
    P_{\ac} &=v\,F_{\ac}.
\end{align}
To make a rough estimation, we consider the supersonic regime $v\gg c_s$, neglecting the dependence of a sound speed in the $F_{\DF}$ and $F_{\ac}$.
We also assume that $\rho_\halo$ is a constant.
With using $v=r \omegas$ and Eq.~\eqref{eq:omegas}, we can rewrite Eq.~\eqref{eq:eng_con} as
\begin{equation}\label{eq:eng_con_dim}
    -\frac{\dif y}{\dif x} = A_\GW\ y^{11/3} +A_\DF + A_\ac,
\end{equation}
with
\begin{align}
     A_\GW &=\frac{2^{7/6}q}{15\sqrt{3}(1+q)^{5/3}},\\
      A_\DF &=162 \times 2^{1/3}\pi \ q(1+q)^{5/3}\ \frac{\rSch^3\ \rho_\halo}{\m1} I,\\
         A_\ac &=162 \times 2^{1/3}\pi \ q(1+q)^{-1/3}\ \frac{\rSch^3\ \rho_\halo}{\m1} .
\end{align}
Here the dimensionless quantities $y$ and $x$ are defined as
\begin{align}
    y &\equiv \frac{\omega_\GW}{\omega_{\GW,\ISCO}} ,\\
    x &\equiv \tau\,\omega_{\GW,\ISCO} ,
\end{align}
with
\begin{equation}
    \omega_{\GW,\ISCO}=\frac{c^3}{3\sqrt{6} G M}.
\end{equation}
The difference in the phase of GWs can be written as
\begin{equation}
\begin{aligned}
    &\Delta\Psi (\tau) \\
    &= \int_0^{\tau } \dif \tau' \{ \omega_\GW(\tau';\text{vacuum})-\omega_\GW(\tau';\text{with DM halo}) \}\\
    &= \int_0^{x} \dif x' \{ y(x';\text{vacuum})-y(x';\text{with DM halo}) \} \\
     & \simeq \frac{3(A_\DF+A_\ac)}{19 A_\GW^2} \left\{ \frac{4}{3} A_{\rm GW}^2 x^2 + A_{\rm GW} x -1 \right.\\
     &\left.~~~~~~~~~~~~~~~~~~~~~~~~~~~~~~~
     + \left[ (1 + \frac{8}{3} A_{\rm GW} x) \right]^{-3/8}  \right\}.
\end{aligned}
\label{eq:analytical}
\end{equation}

The LISA is in a heliocentric orbit and consists of an equilateral triangle formed by three spacecrafts, each separated by a distance of 2.5 million kilometers from one another.
The center of mass for the constellation, known as the guiding center, is in a circular orbit at $1~\AU$ and $20^{\circ}$ behind the Earth.
Choosing the polar coordinate system with the Sun at its origin, the strain of GWs in a detector is given by~\cite{Rubbo:2003ap}
\begin{equation}
    \begin{aligned}
        h(t)=&h_{+}(t-\Delta t)F_{+}(\vartheta,\varphi,\chi,t-\Delta t) \\
        &+ h_{\times}(t-\Delta t)F_{\times}(\vartheta,\varphi,\chi,t-\Delta t),
    \end{aligned}
\end{equation}
where $F_{+}$ and $F_{\times}$ are the detector response functions, $\vartheta$ and $\varphi$ are the latitude
and longitude of the binary in the polar coordinate system, and $\chi$ is the polarization angle.
Here, $\Delta t$ is the delay between the arrival time of GWs at the Sun and the arrival time at the detector, which is given by
\begin{equation}
        \Delta t = -\frac{1\AU}{c}\sin\vartheta\cos(\alpha-\varphi).
\end{equation}
The detector response functions are given by
\begin{equation}
    F_{+}=\frac{1}{2} \left[ D_{+}\cos 2\chi-D_{\times}\sin 2\chi \right],
\end{equation}
\begin{equation}
    F_{\times}=\frac{1}{2} \left[ D_{+}\sin 2\chi+D_{\times}\cos 2\chi \right],
\end{equation}
with
\begin{equation}
    \begin{aligned}
        D_{+}=&\frac{\sqrt{3} }{64} \{-36 \sin ^2\vartheta ~ \sin (2 \alpha -2 \beta ) \\
              & -4 \sqrt{3} \sin 2 \vartheta [ \sin (3 \alpha -2 \beta -\varphi )-3 \sin (\alpha -2 \beta +\varphi ) ]\\
              &+[ \cos 2 \vartheta +3 ] [\cos 2 \varphi  (9 \sin 2 \beta -\sin (4 \alpha -2 \beta ))\\
              &~~~~~~~~~~~~~~~~~~+\sin 2 \varphi  (\cos (4 \alpha -2 \beta )-9 \cos 2 \beta )] \},
    \end{aligned}
\end{equation}
\begin{equation}
    \begin{aligned}
        D_{\times}= &\frac{1}{16} \{\sqrt{3} \cos \vartheta [9 \cos (2 \beta -2 \varphi )-\cos (4 \alpha -2 \beta -2 \varphi )]\\
                    &-6 \sin \vartheta [\cos (3 \alpha -2 \beta -\varphi )+3 \cos (\alpha -2 \beta +\varphi )]\}.
    \end{aligned}
\end{equation}
Here, $\alpha=2\pi t/\yr+\alpha_{0}$ is the orbital phase of the guiding center, and $\beta=2\pi n/3 +\beta_{0} \;(\text{with }n=0,1,2$ for three spacecrafts) is the relative phase of the spacecraft within the constellation.
The parameters $\alpha_{0}$ and $\beta_{0}$ give the initial ecliptic longitude and orientation of the constellation.

In the limit of a large signal-to-noise ratio (SNR), the posterior probability distribution of the source parameters can be approximated by a multivariate Gaussian distribution centered around the true values.
The corresponding covariance can be estimated by the inverse of the Fisher information matrix.
For a network including $N$ independent detectors, the Fisher matrix can be written as
\begin{equation}\label{eq:Fisher}
    \Gamma_{ij} = \left( \frac{\partial \bm{d} (f)}{\partial \theta_{i}} , \frac{\partial \bm{d} (f)}{\partial \theta_{j}} \right)_{\bm{\theta}=\hat{\bm{\theta}}},
\end{equation}
where $\bm{d}$ is given by
\begin{equation}
    \bm{d}(f)= \left[ \frac{\tilde{h}_{1}(f)}{\sqrt{S_{1}(f)}} , \frac{\tilde{h}_{2}(f)}{\sqrt{S_{2}(f)}} , \dots , \frac{\tilde{h}_{N}(f)}{\sqrt{S_{N}(f)}}\right]^{\mathrm{T}} ,
\end{equation}
and $\bm{\theta}$ denotes the parameter vector with true value $\hat{\bm{\theta}}$.
Here $S_{i}(f)$ is the noise power spectral density for the $i$th detector and  $\tilde{h}_{i}(f)$ is the Fourier transformation of the time domain signal.
The bracket operator $(A,B)$ for two functions $A(t)$ and $B(t)$ is defined as
\begin{equation}
    (A,B) = 2 \int_{f_{\min}}^{f_{\max}} \dif f \left[ \tilde{A}(f)\tilde{B}^{*}(f) + \tilde{A}^{*}(f)\tilde{B}(f) \right],
\end{equation}
where $A^{*}$ is its complex conjugate.
The total SNR is given by $\sqrt{(\bm{d}, \bm{d})}$.

The root-mean-squared ($1\sigma$) errors of parameters and the correlation coefficients among parameters can be estimated by the inverse of the Fisher matrix $\bm{\Sigma}=\bm{\Gamma}^{-1}$, which can be written as
 \begin{equation}
     \sigma_{\theta_{i}} = \sqrt{\Sigma_{ii}},
\end{equation}
and
\begin{equation}
    c_{\theta_{i},\theta_{j}} = \frac{\Sigma_{ij}}{\sigma_{\theta_{i}} \sigma_{\theta_{j}}}.
    \label{corr}
\end{equation}

\begin{figure}[htbp]
    \centering
    \renewcommand{\arraystretch}{0}
    \setlength{\tabcolsep}{0pt}
    \begin{tabular}{ccc}
        \includegraphics[]{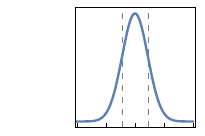}&&\\
        \includegraphics[]{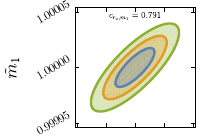}&
        \includegraphics[]{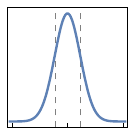}&\\
        \includegraphics[]{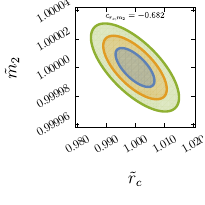}&
        \includegraphics[]{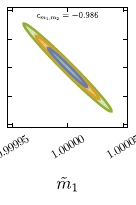}&
        \includegraphics[]{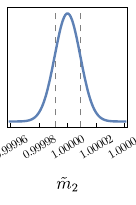}\\
    \end{tabular}
    \caption{Probability distribution of $\{\rc, \m1, \m2\}$ assuming a half year observation with LISA.
    The one-dimension marginalized likelihood and 1-, 2-, 3-$\sigma$ confidence ellipses are shown.
    The fiducial values of parameters are $\{100\rSch, 10^{4}\Msun, 1\Msun\}$, and the gravitational wave SNR is 100.
    }
    \label{fig:corner}
\end{figure}

For the binary with SIDM halo, the parameter vector is $\bm{\theta}=\{\rc; \m1, \m2, D_{L}, \iota, \chi, \vartheta, \varphi, \phi_{\ISCO}, t_{\ISCO} \}$, where $\phi_{\ISCO}$ and $t_{\ISCO}$ are the source phase and time respectively at ISCO.
We consider the binary system with masses given in Table~\ref{tab:paras}, $\{ \iota, \chi, \vartheta, \varphi \}$ are set to be $\pi/4$,  $\{\phi_{\ISCO}, t_{\ISCO}\}$ are set to be $0$, and $D_{L}$ is set to vary the SNR.
Figure~\ref{fig:corner} shows the probability distribution of $\{\rc, \m1, \m2\}$ with the fiducial parameter values $\hat{\bm{\theta}}$ given by set 1 in Table~\ref{tab:paras} and $\rc=100\rSch$.
The tilded parameter is defined as the value with respect to the fiducial value adopted in the Fisher matrix analysis $\tilde{\bm{\theta}}=\bm{\theta}/\hat{\bm{\theta}}$.
The dotted vertical lines in the one-dimensional likelihood function marginalized over the other parameters indicate the $1\sigma$ interval for the parameters.
Two-dimensional ellipses are the 1-, 2-, and 3-$\sigma$ confidence contour plots.
The correlation values defined by Eq.~\eqref{corr} are also shown in the figure and $\rc$ is highly correlated with $\m1$ and anticorrelated with $\m2$.
This correlation arises because the increase in $r_c$ results in the decrease in the soliton density amplitude, and such an effect on the GW signals can be compensated by the increase (decrease) of the central (accompanying) BH mass.

\begin{figure}[htbp]
    \centering
    \includegraphics[]{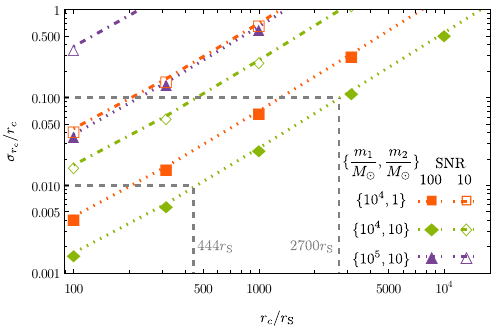}
    \caption{$1\sigma$ uncertainty of the SIDM soliton radius $\sigma_{r_c}$ as a function of $r_c$ ($x$ axis is scaled to the
    Schwarzschild radius and $y$ axis is scaled to a given fiducial value in the Fisher matrix analysis).
    The SNR is set to $100$ and $10$ for illustration and a half year of observation by LISA is assumed.
    To illustrate the soliton size of interest, the gray dotted lines indicate the relative accuracy of 0.1 and 0.01.
    }
    \label{fig:error}
\end{figure}

Focusing on the DM related parameters, the error of $r_c (\propto \sqrt{\lambda}/m^2)$ is shown in \Fig\ref{fig:error}.
The existence of the soliton core is the characteristic feature of SIDM,
and the SIDM parameter space for which the error on $r_c$ is small enough is of our particular interest.
For the parameter range of our interest, the stellar BH is at a radius around ${\cal O}(1 \sim 10) \rSch$ a half year before it reaches the ISCO.
For instance, for $\{m_1/M_{\odot},m_2/M_{\odot}\}=\{10^4,1\}$ illustrated in \Fig\ref{fig:error}, a half year observation of LISA corresponds to receiving the signals when the stellar BH moves though the SIDM soliton from $r\sim 18 \rSch$ to $r\sim 3 \rSch$.
The stellar BH is hence well inside the soliton core during the observation, and a bigger soliton core density leads to a bigger dynamical friction and accretion.
We can hence expect that the signal can be more sensitive to the parameter space which leads to a bigger soliton core density.
Our modeling of the DM profile around a central BH in Sec.~\ref{sec:soliton} has led to a bigger soliton density for a smaller soliton radius and the accuracy is indeed better for a smaller soliton size $r_c$.

Figure~\ref{fig:error} shows that a half year of LISA observation is expected to measure the characteristic radius of soliton $\rc$ within $1\%$ ($10\%$) accuracy when $\rc \lesssim 444\rSch$ ($\rc \lesssim 2700\rSch$).
The binary with a smaller total mass $M$ and a larger mass ratio $q$ increases the expected accuracy for a given $\rc/\rSch$ because it results in a bigger dephasing effect.

\begin{figure}[htbp]
    \centering
    \includegraphics[]{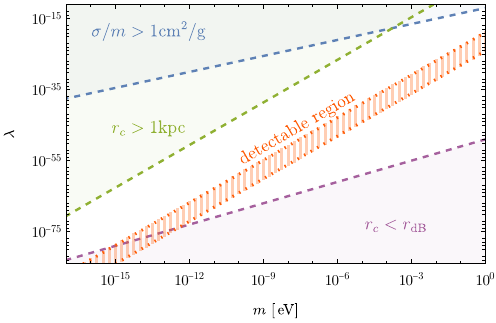}
    \caption{
    The detectable region (orange hatched) in the parameter space of $\{ m,\lambda \}$ from a half year LISA observation of the binary system by requiring $\sigma_{r_c}/r_c<1$.
The dashed purple line corresponds to the limit $\rc \gg \lambda_{\text{dB}}$, and the bullet cluster constraint $\sigma/m \lesssim 1 \rm{ cm}^2/g$ is represented by the dashed blue line.
    The dashed green line represents the Jeans scale bound $r_J \lesssim 1 \text{ kpc}$.
Further information is provided in the main text.}
    \label{fig:lambda-m}
\end{figure}

We can also argue how the black hole masses affect the estimated parameter accuracy illustrated in \Fig\ref{fig:error} by looking at their effects on the dephasing.
The analytical expression of Eq.~(\ref{eq:analytical}) also helps in understanding the behaviors of \Fig\ref{fig:error}.
The difference in dephasing with different values of $\m2$ is partly due to the difference in the dynamical friction which becomes bigger for a bigger $m_2$ when the central IMBH mass is fixed.
The dynamical friction is also sensitive to the velocity of the stellar mass BH and equivalently to the frequency of the orbit.
Such a dependence of the dynamical friction on the frequency is significant as inferred from $v^{-2}$ dependence in Eq.~\eqref{eq:DF_SIDM}.
Consequently, the dephasing for the $m_1=10^5M_{\odot}$ example becomes smaller compared with the $m_1=10^4M_{\odot}$ examples in \Fig\ref{fig:error} because a larger central mass leads to a larger initial orbital radius and slower orbital motion.

Based on the results in \Fig\ref{fig:error}, we can obtain a detectable region (orange hatched) in the parameter space of $\{ m,\lambda \}$ by requiring $\sigma_{r_c}/r_c<1$ in \Fig\ref{fig:lambda-m}.
The dashed purple line corresponds to the limit where the characteristic soliton scale is larger than the de Brogile wavelength $\rc \gg \lambda_{\text{dB}}$ to avoid the effect of quantum pressure~\cite{Boudon:2022dxi}.
The dashed blue line corresponds to the bullet cluster constraint $\sigma/m \lesssim 1 \rm{ cm}^2/g$.
To form enough subgalaxy scale structures, we demand a Jeans length should not exceed 1 kpc scale \, $r_J \lesssim 1 \text{ kpc}$ (dashed green line)~\cite{Goodman:2000tg}.
The Jeans length of the SIDM model with a quartic self-interaction is determined by a characteristic soliton scale $r_J = r_c$ (which is obtained by balancing the self-interacting repulsive force and the gravitational attraction force) ~\cite{Goodman:2000tg, Brax:2019fzb}.

Therefore, our gravitational probes on the SIDM model will be able to shed light on the uncharted area of the parameter space that has not been surveyed by other measurements.
However, the limiting factor of our study is the presence of a degeneracy between two parameters $\lambda$ and $m$ arising from their underlying connection to $r_c (\propto \sqrt{\lambda}/m^2)$.
To break this degeneracy, one would need other probes which possess different dependence on the DM parameters, for instance, the observables which are not directly linked to the height of a soliton profile.
Such an inquiry is left for future work.

\section{Conclusions}\label{sec:dis}

We have studied how SIDM can be probed by forthcoming GW experiments.
It would be ideal to identify a region with a large DM density for effective DM exploration, and the overdense region surrounding a BH provides a promising environment for studying DM properties.
Through the Fisher matrix analysis, we have demonstrated that SIDM scenarios can be distinguished from the vacuum scenario, and we clarified the previously unexplored SIDM parameter space which can be probed by the planned GW experiments.
The future space-borne GW detectors hence would offer a promising means to study and identify DM, which can complement ongoing terrestrial DM search experiments.
While we focused on the LISA specifications in our analysis, it has been proposed that space-based GW detectors, such as LISA and Taiji, can form an observation network~\cite{Ruan:2020smc}.
With the aid of such a network in the future, the observable parameter space can be further enlarged.
The anticipated data from future GW experiments hold great promise in shedding light on the nature of DM and providing insights into the properties of SIDM.

Before concluding our discussions, it is worth mentioning that a more precise estimation of the dynamical friction and accretion processes in the presence of SIDM around a BH deserves further exploration beyond the simplified estimation used in our current modeling.
Additionally, investigating the DM profile of SIDM around a BH, which extends beyond the scope of our modeling presented in this paper, represents an intriguing topic for future research.
For instance, in our study, we employed the Schr\"{o}dinger-Poisson equations to describe the DM profile around the BH, treating it as a source of fixed, external gravitational potential.
This approach provides a reasonable starting point for exploring GW probes, especially considering that the dephasing of the waveform predominantly originates from the radius farther away from the Schwarzschild radius due to the larger number of orbital cycles experienced by the stellar BH.
However, further investigations with more precise treatments, such as utilizing the Einstein-Klein-Gordon equations, are warranted and left for future work.

\begin{acknowledgments}
We thank Wen-Hong Ruan for helpful discussions.
K.~K. thanks the KIAS for hospitality during which this project was initiated.
J.~H.~K. is supported in part by the National Research Foundation of Korea (NRF) grant funded by the Korea government (MSIT) (No. 2021R1C1C1005076), and in part by the international cooperation program managed by the National Research Foundation of Korea (No. 2022K2A9A2A15000153, FY2022).
J.~H.~K. acknowledges hospitality support at APCTP where part of this work was done.
P.~K. is supported in part by the KIAS Individual Grant No.
PG021403, and by National Research Foundation of Korea (NRF) Research Grant NRF-2019R1A2C3005009.
X.~Y.~Y. is supported in part by the KIAS Individual Grant No.
QP090701.
\end{acknowledgments}

\bibliography{citeLib}

\end{document}